# Single-shot 3D characterization the spatiotemporal optical vortex via a spatiotemporal wavefront sensor (STWFS)

Xiuyu Yao[1,2,*], Ping Zhu[1,*], Youjian Yi[1], Zezhao Gong[1,2], Dongjun Zhang[1], Ailin Guo[1], Fucai Ding[1,2], Xiao Liang[1], Xuejie Zhang[1], Meizhi Sun[1], Qiang Zhang[1,2], Miaoyan Tong[1,2], Lijie Cui[1,3], Hailun Zen[1,4], Xinglong Xie[1,*] and Jianqiang Zhu[1,*]

[1]Key Laboratory of High Power Laser and Physics, Shanghai Institute of Optics and Fine Mechanics, Chinese Academy of Sciences, Shanghai 201800, China.
[2]Center of Materials Science and Optoelectronics Engineering, University of Chinese Academy of Sciences, Beijing 100049, China.
[3]School of Physics, Xi'an Jiaotong University, Xi'an 710049, China.
[4]College of Information Science and Technology, Donghua University, Shanghai, China.
These authors contributed equally: Xiuyu Yao, Ping Zhu
* Corresponding authors: zhp1990@siom.ac.cn (P. Zhu); yaoxy@siom.ac.cn (X. Yao); xiexl329@siom.ac.cn (X. Xie); jqzhu@siom.ac.cn (J. Zhu)

**Abstract:** The advent of spatiotemporal wave packets (STWPs), represented by spatiotemporal optical vortices (STOVs), has paved the way for the exploration in optics and photonics. To date, despite considerable efforts, a comprehensive and efficient practical means to characterizing wave packets with such complex structures is still lacking. In this study, we introduced a new method designed to achieve high-precision and high-throughput spatiotemporal wave packet measurements using a user-friendly set up. This method is based on a quadriwave lateral shearing interferometric wavefront sensor that utilizes wavelength division multiplexing, termed the "spatiotemporal wavefront sensor (STWFS)." Using this method, we have fabricated a compact prototype with 295 × 295 spatial pixels × 36 wavelength channels of 0.5 nm spectral resolution in a single frame. This STWFS enabled, for the first time, single-shot self-referenced spatiotemporal three-dimensional (3D) optical field characterizations of STOV pulses with transverse orbital angular momenta $L$ of 1 and 2, and obtained the dynamic visualization of the focused propagation of STOV pulses. Furthermore, the STWFS provides a 1.87-nm (0.95%) root mean square (RMS) absolute accuracy for spatiotemporal phase reconstruction. This achievement represents the highest performance compared with other three-dimensional spatiotemporal metrology methods. As a spatiotemporal optical field characterization method, the STWFS offers ultrafast 3D diagnostics, contributing to spatiotemporal photonics and broader applications across different fields, such as light–matter interactions and optical communications.

## Introduction

Ultrafast light pulses are widely used in various fields [1]. The ability to sculpt structured spatiotemporal wave packets (STWPs) has become increasingly important for revealing the nature of physics and demanding industrial applications [2], including secondary light sources [3] [4] [5] [6], quantum optics [7], stimulated emission depletion microscopy (STED) [8], and simultaneous spatial and temporal focusing [9]. In 2020, the experiments on photonic cyclones carrying a transverse orbital angular momentum (spatiotemporal optical vortices [STOVs]) [10] opened up new avenues for the modulation of STWPs using linear optical methods. These STWPs exhibit new properties that have attracted increasing research interest, including propagation [11], refraction and reflection [12], nonlinear optics [13] [14], tightly focused spin-orbit coupling [15] [16], and metasurface [17] [18] [19]. The unique properties of STOVs have led to important applications in the fields of spatiotemporal differential imaging [20], optical communications [21] and strong field physics [22]. Recently, the ability to sculpt STWPs has moved deeper into higher spatiotemporal dimensions such as complex amplitude and three-dimensionality [23] [24] [25].

Although many exotic spatiotemporal wave packets represented by STOVs have been generated, characterization remains a significant challenge that limits the further development of STWP in the research on light–matter interactions and demanding industrial applications [26]. The most stringent challenge arises from low repetition rate light sources in light-matter interactions, which impose three critical requirements for measurement devices: (1) single-shot, driven by inter-shot fluctuations and the inherently low repetition rate of high-power systems; (2) self-referenced, as stabilizing an additional reference beam is nearly impossible; and (3) sufficiently high 3D accuracy, since only near-field wavepacket information can be observed and modulated in light–matter interaction applications, highlighting the critical importance of measurement precision to deliver desired results. Additionally, rapid feedback capabilities offer significant advantages for the practical implementation of measurement methods. Ideally, achieving single-shot, self-referenced characterization of these three-dimensional (3D) wave packets in the spatiotemporal domain conveniently (common path and in situ), accurately (10-nm root mean square (RMS)), and rapidly (>10 fps) as using wavefront sensors in the spatial domain would make it possible to establish a closed-loop spatiotemporal adaptive control system [26], providing the desired results of light–matter interactions on the target.

Unlike traditional spatiotemporal field measurements that required low-order spatiotemporal coupling, the complex structural information of these spatiotemporal wavepackets is distributed not only in the spatial domain but also in spectral or temporal dimensions. This presents new challenges for measurement systems, which must now possess both precise spatial field resolution and sufficient spectral channels and resolution to capture rapidly changing frequency-domain information.

Currently, methods for achieving 3D measurement of such spatiotemporal structured light fields often rely on mechanical components to perform interferometric scanning in the time or spectral dimension [10, 27]. However, these systems typically suffer from drawbacks such as large size and time-consuming operation, particularly in high-power laser systems with a low repetition rate (< 1 Hz). Moreover, time domain scanning demands temporal chirping of the wavepacket being tested to obtain sufficient temporal resolution. This limitation restricts its application in light-intensity-sensitive scenarios. Compressed short pulses are so short that they cannot be measured directly. The Fourier transform relationship $E(x,y,t) = F^{-1}(E(x,y,\omega))$ between the time and frequency domains highlights the crucial necessity of characterizing the spatial-spectral domain of the pulse. Thus, single shot acquisition of sufficient spectral information in the spatial-spectral domain becomes a feasible solution.

It is extremely difficult to achieve 3D detection with a single shot using a two-dimensional (2D) camera (i.e., to obtain a spectral resolution). One type of compromised approach is to observe only two certain dimensions just like an imaging spectrometer and introduce an additional reference beam to form an interferogram, to obtain phase resolution: the transient-grating supercontinuum spectral interferometer [28] and the spatially resolved spectral interferometer [29]. Although it realizes the single-shot measurement of STOV, it still requires a reference beam, and it is insufficient to observe only two-dimensional features.

We have examined other single-shot 3D spatiotemporal metrology methods, to our knowledge, have not yet demonstrated the ability to measure spatiotemporal singularities or complex structured wavepackets. These include the (COFT) [30] [31] based on coded aperture snapshot spectral imaging (CASSI) and BBSSP based on ptychography [32, 33], encode multispectral phase information onto a 2D camera to achieve single-shot 3D measurement and decode the data cube using compressive sensing or diffraction calculation optimization algorithms. Although these systems have achieved three-dimensional metrology for short pulses, the process of separating overlapping spectra sacrifices considerable reconstruction accuracy and this requires extremely high computational complexity [34], which limits the accurate and high-throughput rapid feedback or further realization of the expected adaptive spatiotemporal light field control [26]. Another category of methods directs both the reference beam and the test beam into a spatial fiber array to achieve spatial resolution in spectral interferometry [35], enabling high spectral domain resolution. However, excessively high costs lead to low spatial resolution, restricting its ability to measure the spatial information of spatiotemporal wavepackets.

To achieve robust 3D measurement of short pulses, a relatively reasonable approach is to obtain spectral resolution through narrow band-pass filters (NBPF). For example, introducing more NBPF to expand the channels of the classic Bayer pattern of red, green, and blue (RGB) color camera and combining it with the robust Hartmann wavefront sensing technology to achieve compact 3D light field measurement. However, the spectral resolution of the Bayer filter on the pixel scale is insufficient, which limits it to measuring only low-order spatiotemporal features. Another STRIPED FISH technique based on NBPF and wavelength-division multiplexing holography can achieve single-shot spatiotemporal full-field characterization through a simple Fourier analysis [36]. The use of a single NBPF of standard optical size represents a straightforward and effective method to attain high spectral resolution and a large channel count. However, the requirement for additional reference light has always limited the widespread application of this technology [37]. The representative state-of-the-art spatiotemporal 3D measurement methods are summarized in Table 1, which lists the key parameters for STWPs measurements.

**Table 1 Comparison of different spatiotemporal 3D measurement methods**

| Method | Single shot? | Self-referenced? | Spatial resolution (pixel) | Spectral resolution (bands/nm) | Computational complexity | Phase residual (prism/grating) |
|---|---|---|---|---|---|---|
| TERMITES [38,39] | No | Yes | 219×201 | 50 / Temporal tens of nm | Moderate | 0.8% |
| FALCON [27] | No | Yes | 35×35 | 9 / 10nm | Low | 6.7% |
| STRIPED FISH [36,40] | Yes | No | - | 30 / 3nm | Low | - |
| Fiber array spectral interferometry [35] | Yes | No | 16 | 300 / 0.31nm | Low | 0.3% |
| MSC HM [41] | Yes | Yes | 93×93 | 4 / 25nm | Moderate | 6.6% |
| BBSSP [32] | Yes | Yes | 326 ×326 | 17 / 10nm | High | - |
| COFT [30] | Yes | Yes | - | 19 / 4.2nm | High | 4.5% |
| **STWFS** | **Yes** | **Yes** | **295×295** | **36 / 0.5nm** | **Low** | **0.3%** |

In this study, we proposed a high-precision and user-friendly single-shot self-referenced pulse $E(x,y,t)$ measurement method based on spectral filtering wavelength-division multiplexing and a quadriwave lateral shearing interference wavefront sensor, called a spatiotemporal wavefront sensor (STWFS). The spatiotemporal phase profile in the spectral domain was accurately reconstructed from a shearing interferogram using wavelength-division multiplexing technology, without requiring an additional reference pulse. The discrete interferogram retrieval process has extremely low algorithmic complexity and high phase accuracy. This is particularly advantageous for inline STWP characterization applications, where efficiency and accuracy are crucial.

Using this method, we successfully achieve in situ measurements of the complete spatiotemporal light field $E(x,y,t)$ of STOV pulses with topological L = 1 and 2. The reconstruction results were found to be in good agreement with the numerical simulations. To the best of our knowledge, this is the first implementation of the single-shot 3D characterization of STOV pulses, including the most simplified design. This advancement is essential for portable applications that provide robustness and long-term stability, notably with the same precision as wavefront sensors. These features highlight the great application potential in spatiotemporal photonics, ultrafast diagnosis, light–matter interactions, and optical communications.

**Results**

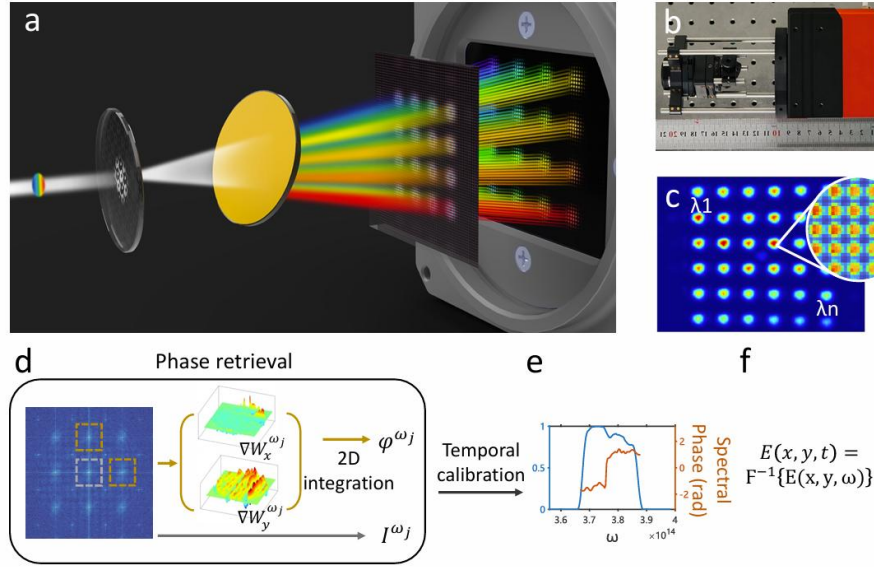

**Fig. 1 Principle of the STWFS system. a,** STWFS system, consisting of a two-dimensional n*n grating, narrow-bandpass filter, and wavefront sensor. Multispectral array of the interferograms on the camera, where each spot is a separate spectral channel. **b,** Photograph of the ultra-compact STWFS developed based on the proposed principle, which is integrated into a camera (20 × 10 × 10 cm in size). **c,** Broadband pulse incident on the STWFS, with an interference pattern formed on the camera. **d,** Using Fourier analysis and two-dimensional phase integration, we extracted spectrally resolved spatial light fields $E_{\text{spatial}}^{\omega}(x,y)$ of the unknown beam. **e,** The last uncertainty term, that is, the spectral phase $\varphi_{x_0,y_0}(\omega)$, is reconstructed from the frequency-resolved optical gating (FROG) trace to help spatial light fields $E_{\text{spatial}}^{\omega}(x,y)$ stitch the phase information. **f,** Inverse Fourier transform is used to retrieve the spatially resolved time light field profile information $E(x,y,t)$.

To obtain spectrum-resolved information in a single frame, the key idea is to map individual spectral channels to different spatial locations within the wavefront sensor using a twisted 2D grating and a narrow band-pass filter (NBPF). This approach avoids any alteration to the structure and accuracy of the sensor. Additionally, the NBPF provide sufficiently high spectral resolution, enabling it to capture high-order and subtle variations between spectral channels. This configuration enables the complete spectrally resolved spatial intensity and phase information ($E_{\text{spatial}}^{\omega}(x,y)$) of the pulse to be captured in a single shot. The architecture of the STWFS system is illustrated in Fig. 1a. It simply consists of two 2D grating, an NBPF, and a camera. This design avoids complex beam manipulation, making it suitable for the characterization of the STWP.

An arbitrary ultrafast optical field is split by 2D grating, which replicates it into multiple subpulses of similar intensities, which are distinct diffraction orders of Mx and My=±1, ±3, and ±5 of the grating. These subpulses are then directed into an NBPF, which is slightly rotated, resulting in a change in the center wavelength of each subpulse owing to the alteration of the incidence angle. These subpulses are approximately monochromatic. The appropriate angle of the NBPF ensures that the spectrum is uniformly dispersed. A wavefront sensor is positioned at a sufficient distance such that the pulse array is completely dispersed, and the 2×2 grating within the wavefront sensor further splits it into four subpulses. These subpulses are spatially sheared, forming an array of interferograms on the camera, as shown in Fig. 2(c). The presented interferogram enables the complete multispectral spatial amplitude profiles $I_{\text{spatial}}^{\omega}(x,y)$ and phase profiles $\varphi_{\text{spatial}}^{\omega}(x,y)$ to be resolved. The interferograms of different wavelengths are spatially independent of one another and can be cropped out to be solved independently for each channel. This setup allows for the use of existing quadriwave lateral shearing wavefront sensors and reconstruction algorithms with appropriate modifications (see Methods), as shown in Fig. 1d. In this study, we used the Fourier transform integration (FTI) method to realize 2D phase retrieval. This process measures the relative phase $\varphi_{\text{spatial}}^{\omega}(x,y)$ of different spatial parts of each frequency, but the phase between the spectra is not captured. Removing this indeterminacy requires measuring the spectral phase $\varphi_{x_0,y_0}(\omega)$ at a given point $(x_0,y_0)$ in space, as shown in Fig. 1e. In our example, this additional measurement was performed using the frequency-resolved optical gating (FROG) at the center of the beam.

Consequently, the 3D data cube $E(x,y,\omega)$ can be reorganized using the wavefront data, and the complete optical field in the time domain $E(x,y,t)$ can be obtained by applying the straightforward Fourier transform as follows:

$$E(x,y,t) = \frac{1}{2\pi}\int E(x,y,\omega)e^{i\omega t}d\omega \tag{1}$$

This measurement configuration provides comprehensive spatiotemporal data of the pulse, including the pulse duration and various spatiotemporal structure variations. Accordingly, the reconstructed light field can be propagated axially, according to the diffraction integral, to obtain the light field information at any position.

*Characterization of the pulse beam carrying transverse orbital angular momentum*

To test the capability of the STWFS, we set up a 4F spatial–spectral $(y,\omega)$ pulse shaper to generate STOV pulses. A commercial Ti:sapphire mode-locked laser system was used as a seed, which generated light with a central wavelength of 800 nm and a collimated Gaussian output with a full-width half maximum (FWHM) of approximately 1 mm. The Gaussian pulses were subjected to spatiotemporal phase shaping $\varphi(y,\omega)$ by a folded 4F system consisting of a grating, cylindrical reflector, and spatial light modulator (SLM). Additional grating pairs compensated for the dispersion to obtain an output near the Fourier transform limit. The output pulse passed through the beam splitter for spectrally resolved spatial light field $E(x,y,\omega)$ measurements, and the reflector arm of the beam splitter BS2 (99:1) entered the FROG for spectral phase $\varphi_{\omega_j}(x_0,y_0)$ measurements. To avoid free-space diffraction from affecting the comparison accuracy of the phase, an achromatic lens was placed in front of the STWFS to create a 2f imaging system to image the SLM plane onto the camera. The camera trigger was synchronized with the SLM clock frequency (60 Hz), and the exposure time required to capture a complete SLM frame ensured stable phase modulation.

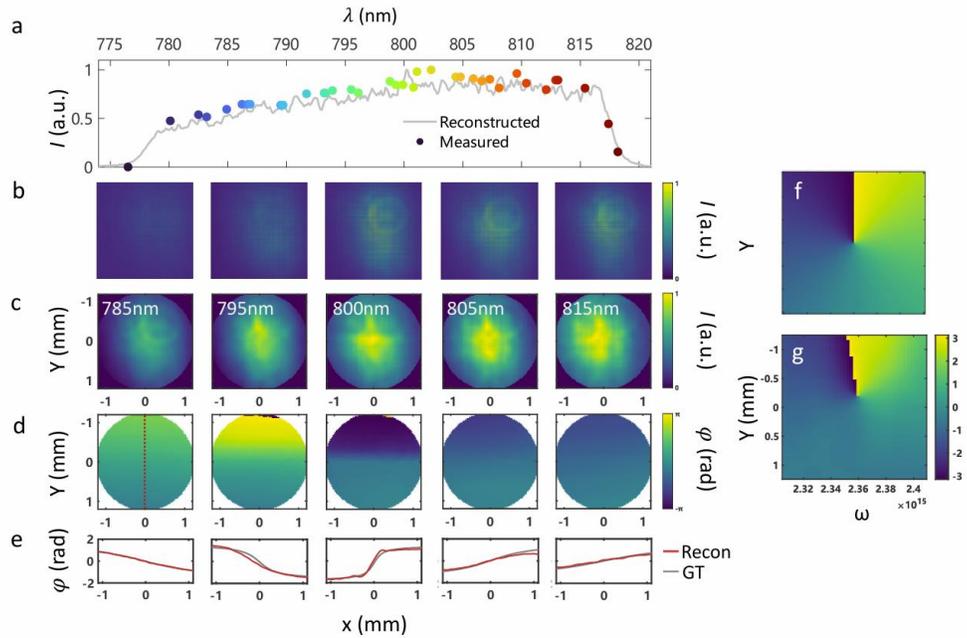

**Fig. 2 Multispectral intensity and phase reconstruction of the l=1 STOV pulse. a**, Comparison of reconstructed (dot) and a fiber-coupled, spectrometer-measured (grey curve) spatially averaged spectrum for the output of the pulse shaping system, with a root mean square error (RMSE) of 3.64%. Selected interferogram (rows **b**), spatial intensity (rows **c**), and phase (rows **d**) of the STOV at five different wavelengths. **e,** Comparison of reconstructed unwrapped phase profiles along x = 0 (Orange curve) and ground truth (grey curve). The reconstructed phase profile along x=0 of 36 spectrum channels (**g**) compared with the phase loaded on the SLM **f**.

The complete light field of the L=1 STOV pulse was measured in the experiment using the STWFS system, as shown in Fig. 2. The 2.6 mm diameter sampling window contained about 120 × 120 spatial sampling points and 36 spectral channels, covering the 776 - 818 nm spectrum, with a spatiotemporal reconstruction speed within 1 second. The interferograms were obtained from the camera and displayed in Fig. 2b for five different frequencies ω (see Supplementary S1 for the complete result of ω). This spatial profile was then used to reconstruct the frequency-resolved 2D spatial intensity, $I_{\omega_j}(x,y)$, as presented in Fig. 2c. Regarding the intensity, the individual spectral channels were not modulated or spatially chirped, as they were not subjected to free-space diffraction. In terms of intensity accuracy, because the NBPF is the only optical element that operates on the spectrum, accurate intensity reconstruction was straightforward. The spatial spectrum was measured using a fiber-coupled spectrometer and subsequently contrasted with the spatial spectrum reconstructed via the STWFS, with a root mean square error (RMSE) of 3.64%, as shown in Fig. 2a. Although the STWFS provides more spatially resolved spectral intensities, it offers a simple verification of the STWFS intensity measurement. The frequency-resolved 2D spatial phases $\varphi_{\omega_j}(x,y)$ were obtained from the interferogram, as displayed in Fig. 2d. The distinguishing feature of the phase was the angular dispersion–like phase shift in the y-direction, with the strongest phase shift pi near the center wavelength. The direction of the phase shift reversed at the central wavelength, which led to a phase singularity in the spatial-frequency domain, as shown in Fig. 2g, whereas the x-direction was barely modulated. In terms of phase accuracy, Fig. 2e illustrates the phase profile in the y-direction (x=0) compared with the ground truth, which was found to be in agreement with the expected results. The results of the phase profile $(y,\omega)$ for the complete 36 spectral channels, as shown in Fig. 2g, were very close to the ground truth (Fig. 2f), which exhibited an orbital angular momentum topology of 1.

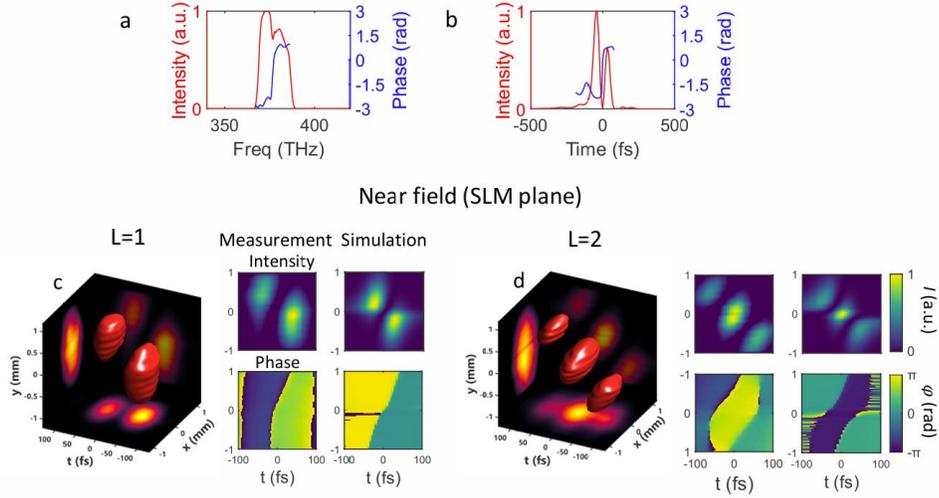

**Fig. 3 Complete 3D reconstructions of the near field profile of an STOV pulse with a spatiotemporal topological charge of L=1,2. (a, b)** Temporal and spectral information of the STOV pulse retrieved from the FROG. **(c, d)** Reconstructed STOV L=1,2 near-field (SLM plane) wave packet 3D profile in the isosurface with a peak intensity of approximately 80%. The side panels show different projections of this intensity, obtained by integrating $|E(x,y,t)|$ along one of the three coordinates. The corresponding intensity and phase profiles along x = 0 are also shown. The near-field spatial intensity $I(x,y)$ is not modulated, but the time-domain profile is stretched, and the number of multiple lobes is equal to L+1. The corresponding simulation results are shown in side panels and Supplementary S2.

To retrieve the STOV pulse in three dimensions, the time domain information at the center of the pulse $(x_0, y_0)$ was measured using the FROG. Of note, the sampling point $(x_0, y_0)$ must be appropriately adjusted to achieve complete spectral coverage, thus enabling the full stitching of the spectral phase $\varphi_{x_0,y_0}(\omega)$ in a single shot. The reconstructed spectral phase $\varphi_{x_0,y_0}(\omega)$ and temporal phase had a 0-pi phase shift at the center wavelength, as shown in Figs. 3a and 3b, with time-domain profiles of a bimodal structure and a pulse duration of 120-fs FWHM. The spectral phase at the sampling point was used to stitch the spatial phase, which was initially solved using the STWFS. Therefore, a simple inverse Fourier transform was applied to reconstruct the near-field 3D spatiotemporal light field $E(x,y,t)$ of the STOV pulse, as shown in Fig. 3c. In the experiment, the complete light field of the STOV pulse with topological L = 2 was reconstructed, as shown in Fig. 3d. To render the structure more visible, the isosurface was set to 80% of the peak intensity. We observed that the energy split in the time domain, NF, and that the number of strings were equal to L+1, which is consistent with the experimental phenomenon [29]. Interestingly, the time-domain structure was generated immediately after the superposition of the spatiotemporal orbital angular momentum, even without spatial–spectral intensity modulation. This cannot be achieved using spectral integration measurement devices, demonstrating the importance of complete spatiotemporal optical field measurements.

STWFS can also observe ultrafast 3D spatiotemporal wavepacket at the microscale. The 3D spatiotemporal field $E(x,y,t)$ in the focal plane (f=150 mm) was obtained by mounting a 10× microscope objective in front of STWFS. The resulting intensity profiles of the pulse are displayed in Figs. 4b and 4e, with side panels showing the optical field $E(y,t)$ at x=0. For L = 1, the spatiotemporal topology of the STOV pulse far-field intensity exhibited a donut structure, with the residual group delay dispersion (GDD) causing a slight splitting of the two flaps [13]. The focal spot was stretched along the y-direction, which is consistent with the simulation results. Furthermore, the phase envelope of 0–2 pi indicates an orbital angular momentum of =1. For L = 2, the far-field intensity exhibited a two-hole structure, which had been observed experimentally [10], and the phase circulations were maintained [42]. The focal spot displayed further stretching in the y-direction compared with that in L=1. Numerical simulations were performed to prove the accuracy of the experimental results (see Supplementary S2 for more details).

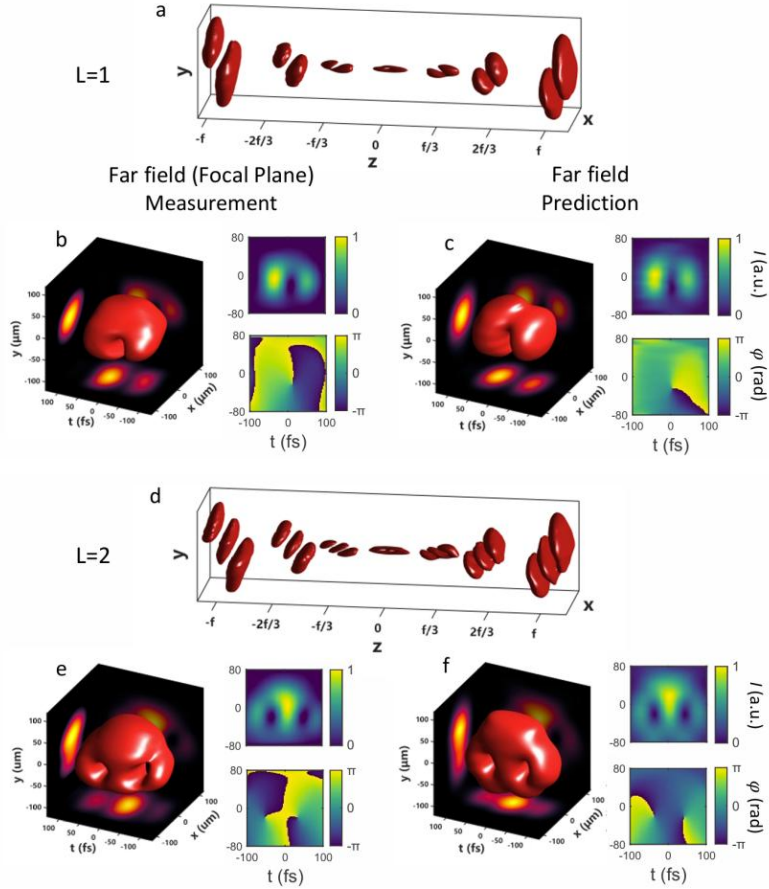

**Fig. 4 Measured and predicted 3D Topological Evolution of STOV During Focusing with a spatiotemporal topological charge of L=1,2. (a, d)** Reconstructed map of intensity topology changes during focusing using spatiotemporal diffraction propagation from z=-f (Figs. 3c and 3d). **(b, e)** Measured far-field (focal plane) wave packet 3D profiles of STOV L=1,2. There is one (L=1), two (L=2) intensities and phase singularities in the spatiotemporal domain $(y, t)$. The far-field spatial intensity $I(x, y)$ is stretched along the y-direction. **(c, f)** Predicted far-field (focal plane) wave packet 3D profiles from Figs. 3c and 3d. Compared to the measured focus **b** and **e**, the wavepackets exhibit highly consistent focal size, topological structure, and intensity distribution of the optical field slices. The RMSE of the 3D intensity distributions is 5% (L=1) and 7% (L=2) within a 1% intensity threshold. The corresponding simulation results are shown in side panels and Supplementary S2.

The 3D spatiotemporal field information of the wavepacket has been fully recorded, enabling prediction of the axial 3D optical field at any position. This capability is particularly valuable for applications such as light-matter interactions or adaptive optics, where direct observation of the focus is challenging. The evolution of the STOV's 3D spatiotemporal field $E(x, y, t)$ during focusing was obtained via diffraction propagation, with topological evolution visualized in Figs. 4a and 4d (see Video Visualization 1, Visualization 2 for detailed dynamics). Near the Rayleigh range, the wavepacket evolves into a donut structure, with the spatiotemporal domain exhibiting a vortex phase. The predicted focal intensity profiles and optical field slices are shown in Figs. 4c and 4f. Compared to the measured focus (Figs. 4b and 4e), the wavepackets are observed to exhibit highly consistent focal size, topological structure, and intensity distribution of the optical field slices. Moreover, the separation of far field spatiotemporal phase singularities at L=2 was accurately predicted. Two-dimensional intensity slices clearly reveal that STWFS accurately predicts slight focal splitting (L=1) and singularity shifts (L=2). We employed the root mean square error (RMSE) as the criterion to evaluate the error between two 3D wavepackets. Calculated within a window of 1/100 normalized intensity, the RMSE values for the predicted and directly measured far field wave packet 3D intensities are 5% for L=1 and 7% for L=2. These mutually corroborated results demonstrate the accuracy of STWFS optical field measurements.

*Accuracy verification test of the STWFS*

We performed experimental tests to verify the precision boundary of the STWFS. The measurement precision of a spectrally resolved spatial light field is vital for the accuracy of spatiotemporal wave packet measurements. A prism can generate angular dispersion, which will result in a pulse front tilt in time domain. The quantity of angular dispersion can be precisely calculated using geometric optics. Moreover, the prism only introduces minute transmission aberrations. Hence, it is highly appropriate to evaluate the precision of spatiotemporal measurement equipment.

We used a Ti:sapphire oscillator with a central wavelength of 800 nm as the input and measured the angular dispersion pulse generated by an equilateral prism (H-ZF13) to test the spatiotemporal phase accuracy of the STWFS, as shown in Fig. 6. To focus on the prism, in this study, the aberrations of the oscillator were subtracted from the phase measurement results ($\varphi_{\text{meas}} = \varphi_{\text{prism}} - \varphi_{\text{osc}}$). Specifically, we first acquired a reference interferogram without the prism under the same optical path. After introducing the prism, we realigned the STWFS and captured the dispersion interferogram, calculating only the incremental dispersion. This approach effectively eliminates errors originating from the incident light's intrinsic dispersion. Five channels of the results are shown in Fig. 5a (see Supplementary S3 for the complete result of ω). Fig. 5b shows the frequency-resolved Zernike coefficient diagram (first 10 terms). We transformed the frequency resolved Zernike coefficient tilt term $C_1^1(\omega)$ to the dispersion angle $\theta$ for comparison with the theoretical value. The angular dispersion $\frac{dD}{d\lambda}$ of light incident on the prism was obtained from the dispersion of the prism material $\frac{dn}{d\lambda}$ as follows:

$$\frac{dD}{d\lambda} = \frac{\cos I_2 \tan I_1' + \sin I_2}{\cos I_2'} \frac{dn}{d\lambda} \qquad (2)$$

where $I_1'$ is the angle between the outgoing normal of the first surface, $I_2$ is the angle of incidence normal of the second surface, and $I_2'$ is the angle of the outgoing normal of the second surface. The angle of incidence was set to $62°\pm 0.14°$. The uncertainty originated from the absolute accuracy of the stepper motor displacement stage, giving rise to an uncertainty of approximately ±0.6 μrad/nm. The angular dispersion curve of the prism was obtained from the material dispersion equation, as shown in Fig. 5c. To comprehensively evaluate the accuracy of individual channels of the STWFS, we calculated the root mean square error (RMSE) between the STWFS - reconstructed dispersion of each channel and the theoretical values. The RMSE turned out to be 0.95%. At the central wavelength, the theoretical angular dispersion value was 130.5 ± 0.6 μrad/nm, while the data reconstructed by STWFS gave a fitted angular dispersion of 130.9 μrad/nm. This led to a 0.3% error, which was in excellent agreement with the theoretical estimates. To the best of our knowledge, this represents the highest phase accuracy among the reported single shot methods.

To further evaluate the precision boundary of the STWFS system, we referred to the precision evaluation standard of the interferometer—the wavefront RMS differential—to verify the absolute accuracy of the spatiotemporal phase. First, we removed the Zernike tilt terms of each channel and then calculated the average wavefronts of the odd-numbered channels $W_{odd}$ and even-numbered channels $W_{even}$ separately. The residual between the two wavefronts was calculated to obtain the absolute accuracy, resulting in a 12.32 nm peak-to-valley (PV) and a 1.87 nm RMS. The residuals are shown in Fig. 5d. These results indicate that the STWFS achieves high spectral resolution while maintaining wavefront sensing precision comparable to common interferometers, with no additional phase accuracy loss, demonstrating its flexibility.

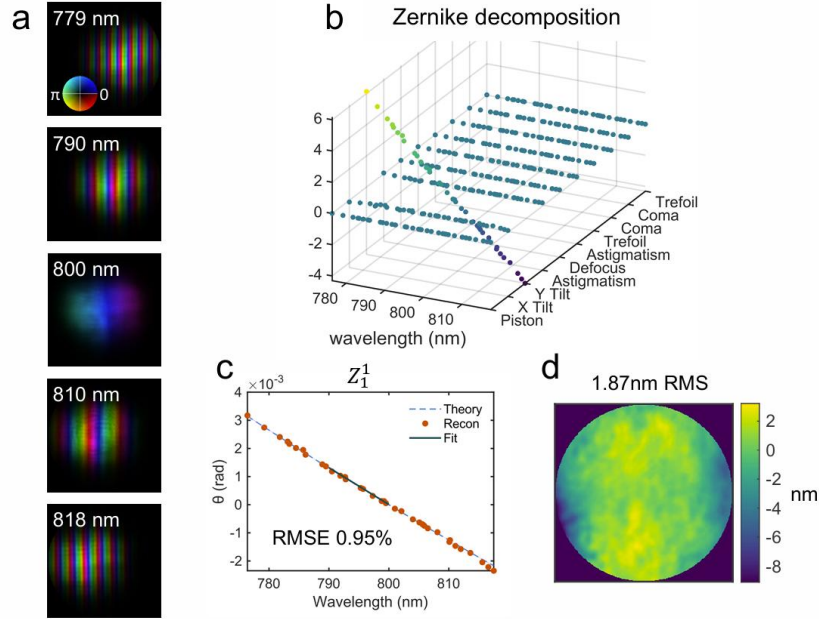

**Fig. 5. Reconstruction of the angular dispersion.** Column **a**, selected five spectrally resolved channels. The amplitude is represented by brightness, and the phase is represented by color. **b,** Frequency-resolved Zernike decomposition of the top 10 terms. Item $z_1^1$ has a clear chromatic aberration. **c,** the reconstruction is compared with the theoretical wavefront tilt angle. The RMSE between the reconstructed and theoretical tilt angles is 0.95%. **d,** Spatiotemporal phase absolute accuracy measurement residual.

## Discussion

The STWFS technique enables straightforward, efficient, and user-friendly acquisition of 3D optical field information for STWPs with ultrahigh spatiotemporal resolution and bandwidth product. However, STWFS has some limitations on the input pulse. On the one hand, spatial multiplexing imposes a trade-off between the number of channels and the field of view (FOV), inherently limited by the camera sensor size. Our prototype supports 36 channels, each with an FOV of approximately 6.6 mm. For applications requiring higher-frequency domain observations, the number of beam-splitting grating points can be increased at the expense of FOV, with careful attention paid to potential sampling overlap between extracted spectral bands. Larger sensors represent an effective solution, and STWFS supports sensor tiling due to its independent channel architecture (detailed in Supplementary S4). On the other hand, this method is limited by polarization because the quadriwave lateral shearing wavefront sensor is fundamentally an interferometric approach and is, therefore, insensitive to polarization applications. A potential solution could involve the use of a polarization camera or two-channel setup. Furthermore, the STWFS supports the measurement of arbitrary polarizations or partially coherent light sources such as light-emitting diodes or halogen lamps.

In summary, we have demonstrated that the STWFS system is capable of measuring ultrafast STWPs with high precision, excellent resolution, scalability, high speed, and a user-friendly setup. By applying the spectral resolution capabilities of wavelength division multiplexing technology within wavefront sensors, the proposed single-shot, self-referenced 3D spatiotemporal technique outperforms current state of the art methods. It addresses critical limitations present in prior approaches through its advancements in both the spatial and spectral domains. Additionally, its rapid and accurate reconstruction capabilities, combined with a user-friendly configuration, render this 3D ultrafast characterization technology feasible for use beyond the laboratory environment.

In the experiment, we demonstrated the reconstruction of STWPs carrying transverse orbital angular momenta L of 1 and 2, which involved multispectral spatial light field reconstruction based on wavelength-division multiplexing wavefront sensing and temporal-domain spectral phase reconstruction based on FROG. Benefiting from the high spectral resolution (0.5nm) and spatiotemporal bandwidth product (2.8 million points) of this method, it enabled us to achieve more accurate and detailed reconstructions, providing deeper insights into the complex light field behaviors, the benefits brought about by the high spectral resolution are intuitively demonstrated in Supplementary S4. We have demonstrated the topological evolution of the 3D focusing process of STOV and verified the ability to precisely predict the focal spots of singular structures. This provides useful references for the field of spatiotemporal photonics and for future applications in precisely controlling the generation of STWPs. In the precision verification experiment, we induced the pulse front tilt by using a prism to verify the boundary of the phase reconstruction precision of the STWFS system, achieving an absolute precision of 1.87-nm RMS in spatiotemporal phase reconstruction. This precision is comparable to that of an interferometer and without additional precision degradation, highlighting the significant advantage of the method in terms of phase precision. This feature stems from the non-interfering and independently reconstructed interference patterns of each channel. Compared to other single-shot 3D optical field metrology methods in Table 1, STWFS avoids accuracy degradation due to inter-channel crosstalk. High precision is of particular significance for applications requiring optical field focusing.

We leveraged STOV pulses as a proof of concept to demonstrate the performance of the STWFS system; however, the utilization of the STWFS has the potential to open novel avenues, such as common path and high-speed features, allowing for the development of a closed-loop adaptive system [26]. Once the high-power spatiotemporal control system reaches maturity, the STWFS can simultaneously detect additional spatiotemporal aberrations in real time, enabling the optimization of spatiotemporal wave packets or peak power at the target, thereby facilitating the desired advanced control of light–matter interactions [14,22].

Furthermore, the STWFS provides spectral resolution while maintaining the existing configuration and accuracy of the quadriwave lateral shearing wavefront sensor. Wavefront sensors have numerous applications in the field of quantitative phase imaging. The STWFS can bring this robust technique into the ultrafast domain. For example, a compact STWFS plays an important role in the real-time observation of events that are difficult to reproduce or nonrepetitive (probabilistic or complex). In particular, the STWFS provides quantitative optical field phase or optical path difference (OPD) information of ultrafast events in a single shot, which has specific applications in biodynamics [43, 44] and laser-induced ultrafast dynamics, such as plasmas, shockwaves, and phase transitions [45-48].

## Materials and methods

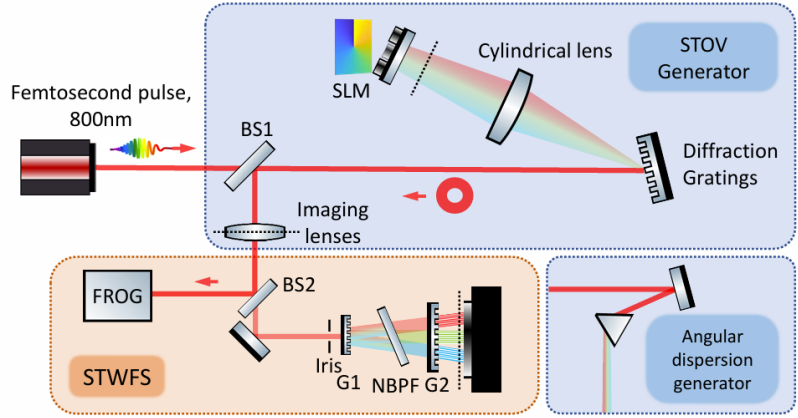

**Fig. 6 System layout to characterize the STOV and angular dispersion pulse.** The STOV pulse is generated by a folded 4F pulse shaper. The STOV pulse enters the STWFS and is divided into two beams by a beam splitter. The reflected beam enters the FROG for spectral phase $\varphi_{x_0,y_0}(\omega)$ measurement in the center of the beam, and the transmitted light enters the multispectral wavefront sensor for spectrally resolved spatial light field $E_{spatial}^{\omega}(x,y)$ measurement. In the accuracy verification experiment, the STOV generator is replaced by a PFT generator to generate the pulse front tilt.

The setup for measuring the STOV is shown in Fig. 6. The ultrashort pulse originated from the oscillator (800-nm central wavelength, 75-MHz repetition frequency, p polarization, 20-fs pulse duration) and entered the 4F spatial–spectral $(y, \omega)$ pulse shaper to generate the STOV. This setup included a blazed grating (1200 lines/mm, gold coating, with an incident angle of approximately 15°), a cylindrical lens (focal length of 100 mm), and a liquid crystal spatial light modulator (2048 × 2048 resolution, 6-μm pixel size, 60 Hz), on which the vortex phase is displayed.

The STWFS system includes an iris to limit the aperture of the pulse, a customized 2D grating (fused silica, 6 × 6 2D beam splitting, period of 21 μm), and a narrow bandpass filter (central wavelength of 830 nm and an FWHM of 0.5 nm) placed close to the 6 × 6 grating. A customized quadriwave 2D grating (fused silica substrate with a period of 45.12 μm) was positioned 2 mm in front of a complementary metal-oxide semiconductor (CMOS) camera (resolution of 14208 × 10640, pixel size of 3.76 μm, and 12-bit monochromatic). The customized single shot FROG included a type-I phase matching crystal with a thickness of 10 um.

### *Algorithm*

The presented interferogram enables the complete multispectral spatial light field $E(x, y, \omega)$ to be resolved. The interferograms of different wavelengths are spatially independent of one another and can be cropped out to be solved independently for each channel. This allows the use of existing quadriwave lateral shearing wavefront sensors and reconstruction algorithms with appropriate modifications. The modifications arise due to the grating and the NBPF. One such modification is that the position of the spot on the camera corresponds to the respective wavelength. Consequently, it is necessary to solve for each channel based on the Apriori wavelength information. Second, according to the angular spectrum diffraction theory, each diffraction order light field $E_{Mx,My}$ undergoes a superposition with different phase shifts $e^{i2\pi(\frac{\alpha_{Mx}}{\lambda}x+\frac{\beta_{My}}{\lambda}y)}$ through the first 2D grating. These phase shifts represent the direction of each diffraction order $(M_x, M_y)$ and are related to the wavelength. After filtering by the NBPF, the discrete spatial wavefronts of each spectral channel are observed to include an additional wavefront tilt $W_{\text{tilt}}^{\omega}(x, y)$. This tilt can be removed by calibrating based on the grating parameters. However, empirical evidence suggests that calibration using a broadband standard light source may yield more optimal results.

The complete wave packet reconstruction process is as follows: the initial step was to crop the interferograms into a multispectral array, with the center determined for the aberration-free beam. The corresponding frequency $\omega$ was provided by the spectrometer under a reference pulse. Subsequently, wavefront extraction was conducted individually for each interferogram, followed by a 2D Fourier transform. The appropriate window size was selected to encompass the dc component $U0_\omega$ and the first-order frequencies in both the x- and y-directions $Ux_\omega$ and $Uy_\omega$, respectively. The light field intensity $I(x,y,\omega)$ was obtained through a 2D inverse Fourier transform of the cropped dc component $U0_\omega$. The inverse Fourier transform was then applied to the cropped and recentered first-order frequency to obtain the wavefront gradients $\nabla_x W_\omega$ and $\nabla_y W_\omega$ along the x- and y- directions, respectively. To eliminate the additional wavefront tilt $W_{\text{tilt}}^{\omega}(x, y)$ brought about by the first grating, the same extraction operation described above is performed on the pre-captured reference map without spatiotemporal aberrations, and subtracting the reference phase gradient before integration gives a more accurate reconstruction result:

$$\nabla_x W_\omega = \frac{d}{4\pi z} \cdot \arg(F^{-1}\{Ux_\omega\} \cdot F^{-1}\{Urefx_\omega\}^*) \qquad (3)$$

$$\nabla_y W_\omega = \frac{d}{4\pi z} \cdot \arg(F^{-1}\{Uy_\omega\} \cdot F^{-1}\{Urefy_\omega\}^*) \qquad (4)$$

where $F^{-1}$ represents the inverse Fourier transform, $d$ is the period of the grating, $z$ is the distance between the grating and the CMOS, and $Urefx_\omega$ is the first order frequency cropped under the Fourier domain of the reference light.

Subsequently, the gradients $\nabla_x W_\omega$ and $\nabla_y W_\omega$ in the two orthogonal directions were integrated in two dimensions to obtain the measured wavefront $W_\omega$. This involves retrieving the 2D scalar field from a 2D vector gradient field. In this study, we reconstructed the spatial phases using the FTI method [49] to limit error propagation, which considers both computational speed and accuracy. Additionally, we utilized Bon's expansion method to mitigate the influence of the Fourier Transform method [50], thereby minimizing the effect of the boundary error. In practice, because the sampling windows and fringe carrier frequencies are identical for each channel in our system, we are able to stack the interferograms into a three-dimensional matrix $(x, y, \lambda)$, which effectively accelerates the reconstruction process.

This process measures the relative phase of different spatial parts of each frequency; however, the phase between the spectra is missing. Removing this indeterminacy requires measuring the spectral phase $\varphi_{x_0,y_0}(\omega)$ at a given point $(x_0, y_0)$ in space. In our example, this additional measurement was performed using the FROG at the center of the beam. Thus, the 3D phase profile was corrected to

$$\varphi(x,y,\omega) = \frac{2\pi}{\lambda} W_\omega(x,y) - \frac{2\pi}{\lambda} W_\omega(x_0,y_0) + \varphi_{x_0,y_0}(\omega) \qquad (5)$$

Consequently, the 3D data cube $E(x, y, \omega)$ can be reorganized using the wavefront data, and the complete optical field in the time domain $E(x, y, t)$ can be obtained by utilizing the straightforward Fourier transform as follows:

$$E(x,y,t) = \frac{1}{2\pi} \int \sqrt{I(x,y,\omega)} e^{i\varphi(x,y,\omega)} e^{i\omega t} d\omega \qquad (6)$$

This measurement configuration provides comprehensive spatiotemporal data of the pulse, including the pulse duration and various spatiotemporal structure variations. Accordingly, the reconstructed light field can be transmitted axially using the diffraction integral to obtain the light field information at any desired position.

## Acknowledgements

This work was supported by the Chinese Academy of Sciences (XDA25020105); National Natural Science Foundation of China (12004403, 12074399, 12204500); Science and Technology Commission of Shanghai Municipality (22YF1455300); Ministry of Science and Technology (2021YFE0116700, 19HJ298801);

## Data availability

The data that support the plots within this paper are available from the corresponding author upon reasonable request. Source data are provided with this paper.

## Code availability

The calculations can be done following the instructions in Methods and Supplementary Information. The code related to this study is available upon reasonable request.

## Author contributions

X. Y., P. Z. and X. X. conceived the idea. X. Y., Y. Y., D. Z., and X. L. performed the theoretical analysis and simulations. X. Y., Z. G., A. G., and F. D. performed the experimental measurements. X. Y., X. Z., M.S., Q. Z., M. T., L. C., H. Z., and J. Z. analyzed the data. All the authors discussed the results and contributed to the manuscript.

## Competing interests

The authors declare no competing interests.